\documentstyle{lamuphys}
\makeatletter
\let\chapter\hid@chapter
\makeatother
\begin{document}
\pagenumbering{arabic}
\title{The radio properties of radio-quiet quasars}

\author{Marek J.\,Kukula\inst{1}, James S.\,Dunlop\inst{1}, David
H.\,Hughes\inst{1} and Steve\, Rawlings\inst{2}}

\institute{Institute for Astronomy, University of Edinburgh,
Royal Observatory, Edinburgh EH9 3HJ, UK
\and
Department of Astrophysics, Nuclear and Astrophysics Laboratory,
University of Oxford, Keble Rd, Oxford OX1 3RH, UK}

\authorrunning{Kukula {\it et al.}}
\maketitle

\begin{abstract}

Although radio-quiet quasars (RQQs), which constitute the majority of
optically-identified quasar samples, are by no means radio silent the
properties of their radio emission are only poorly understood. We
present the results of a multi-frequency VLA study of 27 low-redshift
RQQs. In general, we find that the properties of the radio sources in
RQQs are consistent with them being weak, small-scale ($\sim 1$~kpc)
jets similar to those observed in nearby Seyfert galaxies. We
conclude that a significant fraction of the radio emission in RQQs is
directly associated with the central engine and is not a result of
stellar processes in the surrounding galaxy. There appears to be no
difference between the radio properties of RQQs in elliptical and disc
galaxies, implying that the relationship between the host galaxy and
the `radio loudness' of the active nucleus is not straightforward.

\end{abstract}

\section{Introduction}

Studies of nearby active galaxies have shown that the radio-loud
objects ({\it ie} radio galaxies) are invariably elliptical systems
whereas the radio-quiets (Seyferts) tend to be spirals, suggesting
that it is some property of gas-rich disc galaxies which inhibits the
formation of large, powerful radio sources.  Two factors have
encouraged the extension of this result to the higher redshifts and
larger nuclear luminosities typical of quasars: the success of Unified
Schemes which link radio galaxies and radio-loud quasars (RLQs) via
beaming effects and viewing angle and the fact that Seyfert 1 nuclei
and radio-quiet quasars (RQQs) form a continuous sequence in terms of
their optical luminosities and have identical emission line
characteristics. Thus we have a picture in which RLQs occur in
elliptical galaxies whilst RQQs are found in discs.

However, recent studies have shown that quasars occur in a wide
variety of environments ({\it eg} Bahcall {\it et al.} 1997). In particular
it seems that {\it not all} RQQs lie in disc systems, with as many as
50\% occurring in elliptical galaxies ({\it eg} V\'{e}ron-Cetty \&
Woltjer 1990). Indeed, there is some evidence that elliptical galaxies
might account for all of the most optically luminous RQQs (Taylor {\it
et al.} 1996). Clearly the simple `radio-loud $\equiv$ elliptical,
radio-quiet $\equiv$ disc' picture can no longer be supported, and it
has become more important than ever to determine in what respects
radio-quiet quasars are different from their radio-loud
counterparts. Unfortunately, the most obvious wavelength regime in
which they differ - the radio - is also the regime in which least is
known about the properties of RQQs. Most radio surveys of
optically-selected quasar samples have lacked the sensitivity to
detect the majority of radio-quiet objects, and have provided only
limited information on their radio structure and spectral index.

To remedy this situation we embarked on a multi-frequency,
high-resolution radio survey of 27 low-redshift ($0.1 \leq z \leq
0.3$), low-luminosity (M$_{V}>-26$) RQQs using the Very Large Array
(VLA). Observations were made in `A' configuration at 1.4, 4.8 and
8.4~GHz, with angular resolutions of 1.4, 0.4 and 0.24$''$
respectively, and radio emission was detected in 75\% of the
quasars. The survey will be discussed in more detail by Kukula {\it et
al.} (1997).

\section{Results of the survey}

A principal goal of the survey was to investigate the origin of the
radio emission in RQQs and to attempt to distinguish between radio
emission produced by stellar processes ({\it eg} circumnuclear starburst
regions) in the body of the host galaxy and emission which is directly
associated with the quasar ({\it eg} radio jets). Our findings can be
summarized as follows:

{\bf (1)} In the majority of objects we detect an unresolved ($\leq
1$~kpc) radio component which is coincident with the optical position
of the quasar.

{\bf (2)} The radio spectral indices of the RQQs are generally steep
($\alpha \sim 0.7$, where $S \propto \nu^{-\alpha}$), although two
objects, which also exhibit unusually large radio luminosities and
optical variability, have flat radio spectra. Following Falcke,
Sherwood \& Patnaik (1996) we interpret these flat-spectrum objects as
RQQs in which a relativistic jet is closely aligned to the line of
sight, leading to doppler boosting of the radio emission.

{\bf (3)} Lower limits on the brightness temperatures of the radio
sources place them at the upper end of the range expected for emission
related to stellar processes ($T_{B} \sim 10^{5}$~K). In many objects
the brightness temperature is almost certainly many orders of
magnitude greater than this, in which case the radio emission {\it
cannot} be produced by a starburst. Observations with greater angular
resolution will be required in order to confirm these results.

{\bf (4)} In five objects we are able to resolve radio structure,
which takes the form of double, triple and linear sources on scales of
a few kiloparsecs. These images resemble early, low-resolution maps of
nearby Seyfert nuclei - objects which have since been shown to contain
small-scale ($\leq 1$~kpc), highly-collimated radio jets.  The current
maps could therefore be taken as evidence that collimated ejection of
radio plasma from the central engine is also occurring in RQQs but
high-resolution VLBI observations will be necessary in order to
confirm this interpretation.

{\bf (5)} The distribution of radio luminosities in the RQQ sample
forms a natural extension to that of Seyfert 1 nuclei. There appears
to be a correlation between radio luminosity and the optical absolute
magnitude of the quasar, implying a close relationship between the
central engine and the mechanism responsible for the bulk of the radio
emission.

We therefore conclude that in the majority of RQQs a significant
fraction of the overall radio emission comes from a compact nuclear
source which is directly associated with the quasar's central
engine. By analogy with Seyfert galaxies this nuclear source probably
takes the form of a small-scale radio jet, qualitatively similar
to the more powerful jets observed in RLQs and radio galaxies.

\section{Radio emission and the hosts of RQQs}

Seventeen of the RQQs in the present sample were also included in our
near-infrared imaging study of quasar hosts (Taylor {\it et al.} 1996;
described elsewhere in this volume by Kukula {\it et al.}). These
observations showed that slightly less than half of the RQQs occur in
galaxies in which the dominant stellar component has a spheroidal
rather than an exponential (disc) distribution.  

We were therefore able to carry out a study of the relationship between
the host morphology and the radio properties of the AGN and to
investigate the extent to which the radio sources in RQQs with
spheroidal hosts could be distinguished from those in discs.  The AGN
traditionally associated with elliptical hosts ({\it ie} radio
galaxies and RLQs) produce large, powerful radio sources and this
suggests that the RQQs in elliptical galaxies, whilst technically
`radio quiet', might still harbour radio sources which differ in size
and luminosity from those in disc galaxies.

However, in our sample we were able to find {\it no} clear separation
between the radio properties of the two types of radio-quiet
quasar. The RQQs with elliptical hosts {\it do not} contain larger or
more luminous radio sources than those in disc galaxies. Both types of
RQQ are equally likely to contain extended radio structure and to show
evidence for collimated radio jets.  The radio sources in elliptical
galaxies show no tendency to have flatter spectra than those in discs
(as might have been expected if the jets in elliptical galaxies are
more likely to be relativistic).

Although the current sample is small, and is limited to RQQs of
relatively low optical luminosity, our radio survey clearly
demonstrates that having an elliptical host does not automatically
confer a large radio luminosity on the active nucleus.  A significant
number of ellipticals with active nuclei are {\it not} producing
large, powerful radio sources but contain small, weak sources which
appear to be identical to those in disc systems. Further, more
detailed studies of the host galaxies will be required in order to
determine if and how these ellipticals differ from those containing
radio loud AGN.

\end{document}